\documentclass[12pt]{article}
\usepackage{times}

\newcommand{\lsim}{\stackrel{<}{_\sim}}
\newcommand{\gsim}{\stackrel{>}{_\sim}}
\usepackage{epsf}
\def\beq{\begin{equation}}
\def\eeq{\end{equation}}
\def\bea{\begin{eqnarray}}
\def\eea{\end{eqnarray}}

\def\nnb{\nonumber}

\def\rar{\rightarrow}
\def\nnb{\nonumber}

\def\ba{\begin{array}}
\def\ea{\end{array}}
\def\bea{\begin{eqnarray}}
\def\eea{\end{eqnarray}}
\def\Bpll{$B_s\rightarrow \,\,\phi\,\, \ell^+ \ell^-$}
\def\Bpmm{$B_s\rightarrow \,\,\phi\,\, \mu^+ \mu^-$}
\def\tep{$b \rar s \ell^+ \ell^-$}

\title{The exclusive \Bpll \,\,decay in the two Higgs doublet models}
\author{\vspace{1cm}\\
         {\bf G\"{u}ray Erkol}
         \thanks{E-mail address:
        gurerk@newton.physics.metu.edu.tr} \, \, and \, \,
        {\bf G\"{u}rsevil  Turan}
        \thanks{E-mail address:
        gsevgur@metu.edu.tr}\,\,\thanks{Common address: Middle East Technical University Physics Dept. Inonu Bul.
06531 Ankara-TURKEY}}
        \date{}
\begin{document}
\setlength{\baselineskip}{24pt} \maketitle
\setlength{\baselineskip}{7mm}

\abstract{We study the differential branching ratio, branching
ratio and the forward-backward asymmetry for the exclusive \Bpll
decay in the two Higgs doublet model. We analyze
the dependencies  of these quantities on the model parameters and
show that these observables are highly sensitive to new physics and hence
may provide powerful probe of the SM and beyond.
%\dedicated{Dedicated to\ldots\\if you want.}
\thispagestyle{empty} \setcounter{page}{1}
\section{Introduction}
The analysis of flavor-changing neutral current (FCNC) decays is one
of the most promising research areas in particle physics from both
theoretical and  experimental sides. The rare B-meson decays induced by
FCNC $b\rightarrow s$ transition,
have received a special attention since their investigation opens up the possibility
of a more precise determination of fundamental parameters of the standard model (SM),
such as the elements of the Cabibbo-Kobayashi-Maskawa (CKM)
matrix, the leptonic decay constants etc.. In addition, the studies on the rare
B-meson decays open a window to investigate the physics beyond the SM, such as
the two Higgs doublet model (2HDM), Minimal Supersymmetric extension of the SM
(MSSM)\cite{Hewet}, etc.  to test these models and make estimates about their free
parameters.

The difficulties present in the experimental investigation of the inclusive decays
stimulate the study of the exclusive decays.  There  exist now upper limits on the branching ratios
of $B^0\rightarrow K^{0*} \mu^+ \mu^-$ and $B^+\rightarrow K^+ \mu^+ \mu^-$, given by
CDF collaboration \cite{Affolder}
\bea
BR (B^0\rightarrow K^{0*} \mu^+ \mu^-) & < & 4.0 \times 10^{-6} \nnb \\
BR (B^+\rightarrow K^+ \mu^+ \mu^-) & < & 5.2 \times 10^{-6} \nnb .
\eea
With these measured upper limits and  also the  recent measurement of the branching ratio of
$B\rightarrow K \ell^+ \ell^-$ with $\ell=e,\mu$,
\bea
BR (B\rightarrow K \ell^+ \ell^-) & = &( 0.75^{+0.25}_{-0.21}\pm 0.09) \times 10^{-6} \nnb ,
\eea
at KEK \cite{Belle}, the processes $B\rightarrow (K,K^*)\ell^+\ell^-$ have received great
interest so that their theoretical calculation  has been the subject of many
investigations in the SM and beyond \cite{dt}-\cite{erkol2}.
Along this line, the exclusive decays
induced by $b\rightarrow s \ell^+\ell^-$ transition, like
$B_s\rightarrow \phi\ell^+\ell^-$, become also attractive since the SM predicts a
relatively larger branching ratio, which is measurable in the near future.
On the other hand, the theoretical analysis of exclusive decays is more complicated and
contains substantial uncertainties due to the hadronic form factors, which need
non-perturbative methods for their calculations. One of the methods that can be used
to calculate the hadronic matrix elements is the three parameter fit of
the light-cone QCD sum rule. In ref.\cite{Ball}, the form factors for
$b\rightarrow s \ell^+\ell^-$ induced exclusive $B_s\rightarrow \phi\ell^+\ell^-$ decay
have been calculated in the framework of this method. More recently, this process
has been also  investigated in a different context, namely using a so-called
Constituent Quark-Meson model (CQM) based on quark-meson interactions \cite{AD}.

In this paper we investigate the exclusive \Bpll decay in the framework of
the 2HDM. The transitions where B-meson decays into a  vector meson,
like $K^*$ \cite{dt}-\cite{erkol2} and $\rho$ \cite{Kruger1}-\cite{Erkol},
have been extensively studied in the literature both in the SM and beyond.
We calculate the dependence of the branching ratio (BR) and the forward backward
asymmetry ($A_{FB}$) of the exclusive \Bpll decay on the model parameters
in the model I, II and III versions of the 2HDM and show that for some values
of these parameters especially model III gives significance contributions to the BR and
$A_{FB}$.

The paper is organized as follows: In section \ref{s2}, after we
present the theoretical framework of the 2HDMs and the leading
order QCD corrected effective Hamiltonian for the process \tep, we
calculate the differential BR and the $A_{FB}$ of the exclusive
\Bpll decay. The \ref{s3}. section is devoted to the numerical
analysis and the discussions.

\section{The theoretical framework}\label{s2}

Before presenting the details of our calculations,  we would like to
summarize the main essential points of the  2HDM, which is one of the most popular
extensions of the SM. The 2HDM has two complex Higgs doublets instead of only one
in the SM. In general, the 2HDM possesses tree-level FCNCs that can be avoided by
imposing an {\it ad hoc} discrete symmetry \cite{Glashow}. As a
result, there appear two different choices, namely model I and II,
depending on whether up-type and down-type quarks couple to the
same or two different Higgs doublets, respectively. Model II has
been more attractive since its Higgs sector is the same as the
Higgs sector in the supersymmetric models. Physical content of the Higgs
sector includes three neutral Higgs bosons $H^0$, $h^0$ and $A^0$, and a pair
of charged Higgs bosons $H^{\pm}$. In these models the interaction vertices of the
Higgs bosons and fermions depend on the ratio $\tan \beta =v_1/v_2$, where
$v_1$ and $v_2$ are the vacuum expectation values of the first and the second Higgs doublet
respectively, and it is a free parameter in the model. The constraints on $\tan \beta $
are usually obtained from $B-\bar{B}$, $K-\bar{K}$ mixing, $b\rightarrow s \, \gamma$
decay width, semileptonic decay $b\rightarrow c \, \tau \bar{\nu}$ and is given by
\cite{ALEP}
\bea
0.7 \leq \tan \beta \leq 0.52 (\frac{m_{H^{\pm}}}{1 ~ GeV}) \, ,
\eea
and the lower bound $m_{H^{\pm}} \geq 200$ GeV has also been given in \cite{ALEP}.

In a more general 2HDM,
namely model III \cite{modelIII,Soni}, no discrete symmetry is
imposed and there appear FCNC naturally at the tree level. We note
that in model III, FCNC receiving contributions from the first two
generations are highly suppressed, which is confirmed by the low
energy experiments. As for those involving the third generation,
it is possible to impose some restrictions on them with the
existing experimental results.  Since the popular models I and II are special cases of
the more general model III, we will give here the details of model III only.

The Yukawa Lagrangian in this general case is written as
\begin{eqnarray}
{\cal{L}}_{Y}&=&\eta^{U}_{ij} \bar{Q}_{i L} \tilde{\psi_{1}} U_{j
R}+ \eta^{D}_{ij} \bar{Q}_{i L} \psi_{1} D_{j R}+ \xi^{U\,
\dagger}_{ij} \bar{Q}_{i L} \tilde{\psi_{2}} U_{j R}+ \xi^{D}_{ij}
\bar{Q}_{i L} \psi_{2} D_{j R} + h.c. \,\,\, , \label{lagrangian}
\end{eqnarray}
where $i,j$  are family indices of quarks , $L$ and $R$ denote
chiral projections $L(R)=1/2(1\mp \gamma_5)$, $\psi_{m}$ for
$m=1,2$, are the two scalar doublets, $Q_{i L}$ are quark
doublets, $U_{j R}$, $D_{j R}$ are the corresponding quark
singlets, $\eta^{U,D}_{ij}$ and $\xi^{U,D}_{ij}$ are the matrices
of the Yukawa couplings.
We can choose two scalar doublets $\varphi_1$ and
$\varphi_2$ in the following form
\begin{eqnarray}
\varphi_{1}=\frac{1}{\sqrt{2}}\left[\left(\begin{array}{c c}
0\\v+H^{0}\end{array}\right)\; + \left(\begin{array}{c c} \sqrt{2}
\chi^{+}\\ i \chi^{0}\end{array}\right) \right]\, ;
\varphi_{2}=\frac{1}{\sqrt{2}}\left(\begin{array}{c c} \sqrt{2} H^{+}\\
H_1+i H_2 \end{array}\right) \,\, \label{choice}
\end{eqnarray}
with the vacuum expectation values,
\begin{eqnarray}
<\varphi_{1}>=\frac{1}{\sqrt{2}}\left(\begin{array}{c c}
0\\v\end{array}\right) \,  \, ; <\varphi_{2}>=0 \,\,
\label{choice2}
\end{eqnarray}
so that the first doublet $\psi_1$ is the same as the one in the
SM, while the second doublet contains all the new particles.
Further, we take $H_1$ and $H_2$ as the mass eigenstates $h^0$ and
$A^0$, respectively.

After the rotation that diagonalizes the quark
mass eigenstates, the part of the Lagrangian that is responsible for the FCNC at the
tree level looks like
\begin{eqnarray}
{\cal{L}}_{Y,FC}= -H^{\dagger} \bar{{\cal{U}}}[V_{CKM}\, \xi^{D}_{N}\,
R-\xi^{U \dagger}_{N}\, V_{CKM} \, L] \,  {\cal{D}}   \,\, ,
\label{lagrangianFC}
\end{eqnarray}
where ${\cal{U}}({\cal{D}})$ represents the mass eigenstates of up (down) type quarks.
In this work, we adopt the following
redefinition of the Yukawa couplings:
\bea \xi^{U,D}_N=\sqrt{\frac{4G_F}{\sqrt{2}}}\bar{\xi}^{U,D}_{N,ij}. \eea

Next step is to calculate the matrix
elements for the inclusive $b\rightarrow s \ell^+ \ell^-$ decay in model III including
the QCD corrections.
For this, the effective Hamiltonian method provides a powerful
framework. The procedure is to match the full theory with the
effective theory, which is obtained by integrating out the heavy
degrees of freedom, i.e., $t$ quark, $W^{\pm}$, $H^{\pm}$, $h^0$
and $H^0$ in our case, at high scale $\mu=m_W$, and then calculate
the Wilson coefficients at the lower scale $\mu \sim {\cal O}(
m_b)$ using the renormalization group equations. Following these
steps above, one can obtain the effective Hamiltonian governing
the $b \rightarrow s \ell^+ \ell^-$ transitions, in model III in
terms of a set of operators
\begin{eqnarray}\label{Hamiltonian}
{\cal H}_{eff} & = & \frac{4 G_F}{\sqrt{2}} \, V_{tb}
V^*_{ts}\Bigg\{ \sum_{i=1}^{10} \, \, C_i (\mu ) \, O_i(\mu) \,
\,\Bigg\}
\end{eqnarray}
Here, $O_1$ and $O_2$ are \emph{the current-current operators},
$O_3$,...,$O_6$ are usually named as \emph{the QCD penguin
operators}, $O_7$ and $O_8$ are \emph{the magnetic penguin
operators} and $O_9$ and $O_{10}$ are \emph{the semileptonic
electroweak penguin operators}. $C_i(\mu)$ are Wilson coefficients
renormalized at the scale $\mu$.
The operator basis in the  2HDM for our process can be found in
\cite{Grinstein,Misiak}.

Denoting the Wilson coefficients for the relevant
process in the SM with $C^{SM}_{i}(m_W)$ and the additional charged Higgs
contributions with $C^{H}_{i}(m_W)$ , we have the initial values given by
\cite{Grinstein,Aliev4}
\begin{eqnarray}
C^{SM}_{1,3,\dots 6}(m_W)&=&0 \nonumber \, \, , \\
C^{SM}_2(m_W)&=&1 \nonumber \, \, , \\
C_7^{SM}(m_W)&=&\frac{3 x_t^3-2 x_t^2}{4(x_t-1)^4} \ln x_t+
\frac{-8 x_t^3-5 x_t^2+7 x_t}{24 (x_t-1)^3} \nonumber \, \, , \\
C_8^{SM}(m_W)&=&-\frac{3 x_t^2}{4(x_t-1)^4} \ln x_t+
\frac{-x_t^3+5 x_t^2+2 x_t}{8 (x_t-1)^3}\nonumber \, \, , \\
C_9^{SM}(m_W)&=&-\frac{1}{sin^2\theta_{W}} B(x_t) + \frac{1-4
\sin^2 \theta_W}{\sin^2 \theta_W} C(x_t)-D(x_t)+
\frac{4}{9}, \nonumber \, \, , \\
C_{10}^{SM}(m_W)&=&\frac{1}{\sin^2\theta_W}
(B(x_t)-C(x_t))\nonumber \,\, ,
\end{eqnarray}
and
\begin{eqnarray}
C^{H}_{1,\dots 6 }(m_W)&=&0 \nonumber \, , \\
C_7^{H}(m_W)&=& Y^2 \, F_{1}(y_t)\, + \, X Y \,  F_{2}(y_t)
\nonumber  \, \, , \\
C_8^{H}(m_W)&=& Y^2 \,  G_{1}(y_t) \, + \, X Y \, G_{2}(y_t)
\nonumber\, \, , \\
C_9^{H}(m_W)&=&  Y^2 \,  H_{1}(y_t) \nonumber  \, \, , \\
C_{10}^{H}(m_W)&=& Y^2 \,  L_{1}(y_t) \label{CH} \, \, ,
\end{eqnarray}
where
\bea
x_t&=&\frac{m_t^2}{m_{W}^2}~~, ~~y_t~=~\frac{m_t^2}{m_{H^\pm}^2}~~, \nnb \\
X & = &
\frac{1}{m_{b}}~~~\left(\bar{\xi}^{D}_{N,bb}+\bar{\xi}^{D}_{N,db}
\frac{V_{td}}{V_{tb}} \right) ~~,~~ \nnb \\
Y & = &
\frac{1}{m_{t}}~~~\left(\bar{\xi}^{U}_{N,tt}+\bar{\xi}^{U}_{N,tc}
\frac{V^{*}_{cd}}{V^{*}_{td}} \right) ~~.
\eea
Note that the results for model I and II can be obtained from
model III by the following substitutions:
\begin{eqnarray}
Y \rightarrow \cot \beta \, \, & , & \,\, X Y\rightarrow -\cot^2
\beta \,\, for\,\,
 model \, I \nnb \\
Y \rightarrow \cot \beta \, \, & , & \,\, X Y\rightarrow 1 \,\,
for\,\, model \, II \, .\nnb
\end{eqnarray}
The explicit forms of the functions $F_{1(2)}(y_t)$,
$G_{1(2)}(y_t)$, $H_{1}(y_t)$ and $L_{1}(y_t)$ in Eq.(\ref{CH})
are given as
\begin{eqnarray}
F_{1}(y_t)&=& \frac{y_t(7-5 y_t-8 y_t^2)}{72 (y_t-1)^3}+
\frac{y_t^2 (3 y_t-2)}{12(y_t-1)^4} \,\ln y_t \nonumber  \,\, ,
\\
F_{2}(y_t)&=& \frac{y_t(5 y_t-3)}{12 (y_t-1)^2}+ \frac{y_t(-3
y_t+2)}{6(y_t-1)^3}\, \ln y_t \nonumber  \,\, ,
\\
G_{1}(y_t)&=& \frac{y_t(-y_t^2+5 y_t+2)}{24 (y_t-1)^3}+
\frac{-y_t^2} {4(y_t-1)^4} \, \ln y_t \nonumber  \,\, ,
\\
G_{2}(y_t)&=& \frac{y_t(y_t-3)}{4 (y_t-1)^2}+\frac{y_t}
{2(y_t-1)^3} \, \ln y_t  \nonumber\,\, ,
\\
H_{1}(y_t)&=& \frac{1-4 sin^2\theta_W}{sin^2\theta_W}\,\,
\frac{xy_t}{8}\,
\left[ \frac{1}{y_t-1}-\frac{1}{(y_t-1)^2} \ln y_t \right]\nonumber \\
&-& y_t \left[\frac{47 y_t^2-79 y_t+38}{108 (y_t-1)^3}- \frac{3
y_t^3-6 y_t+4}{18(y_t-1)^4} \ln y_t \right] \nonumber  \,\, ,
\\
L_{1}(y_t)&=& \frac{1}{sin^2\theta_W} \,\,\frac{x y_t}{8}\,
\left[-\frac{1}{y_t-1}+ \frac{1}{(y_t-1)^2} \ln y_t \right]
\nonumber  \,\, .
\\
\label{F1G1}
\end{eqnarray}
Finally, the initial values of the coefficients in the model III
are
\begin {eqnarray}
C_i^{2HDM}(m_{W})&=&C_i^{SM}(m_{W})+C_i^{H}(m_{W}) \,. \label{CiW}
\end{eqnarray}
Using these initial values, we can calculate the coefficients
$C_{i}^{2HDM}(\mu)$ at any lower scale in the effective theory
with five quarks, namely $u,c,d,s,b$ similar to the SM case
\cite{Grinstein}-\cite{Buras}.

The Wilson  coefficients playing  the essential role in this
process are $C_{7}^{2HDM}(\mu)$, $C_{9}^{2HDM}(\mu)$,
$C_{10}^{2HDM}(\mu)$. For completeness, we also give
their explicit expressions
\begin{eqnarray}
C_{7}^{eff}(\mu)&=&C_{7}^{2HDM}(\mu)+ Q_d \, (C_{5}^{2HDM}(\mu) +
N_c \, C_{6}^{2HDM}(\mu))\nonumber \, \, , \label{C7eff}
\end{eqnarray}
where the leading order QCD corrected Wilson coefficient $C_{7}^{LO,
2HDM}(\mu)$ is given by \cite{Buras,Grinstein,Misiak}
\begin{eqnarray}
C_{7}^{LO, 2HDM}(\mu)&=& \eta^{16/23} C_{7}^{2HDM}(m_{W})+(8/3)
(\eta^{14/23}-\eta^{16/23}) C_{8}^{2HDM}(m_{W})\nonumber \,\, \\
&+& C_{2}^{2HDM}(m_{W}) \sum_{i=1}^{8} h_{i} \eta^{a_{i}} \,\, ,
\label{LOwils}
\end{eqnarray}
and $\eta =\alpha_{s}(m_{W})/\alpha_{s}(\mu)$, $h_{i}$ and $a_{i}$
are the numbers which appear during the evaluation \cite{Buras}.

$C_9^{eff}(\mu)$ contains a perturbative part and a part coming
from LD effects due to conversion of the real $\bar{c}c$ into
lepton pair $\ell^+ \ell^-$:
\begin{eqnarray}
C_9^{eff}(\mu)=C_9^{pert}(\mu)+ Y_{reson}(s)\,\, ,
\label{C9efftot}
\end{eqnarray}
where
\begin{eqnarray}
C_9^{pert}(\mu)&=& C_9^{2HDM}(\mu) \nonumber
\\ &+& h(z,  s) [ 3 C_1(\mu) + C_2(\mu) + 3 C_3(\mu) +
C_4(\mu) + 3 C_5(\mu) + C_6(\mu)] \nonumber \\&-&   \frac{1}{2}
h(1, s) \left( 4 C_3(\mu) + 4 C_4(\mu)
+ 3 C_5(\mu) + C_6(\mu) \right)\nnb \\
&- &  \frac{1}{2} h(0,  s) \left[ C_3(\mu) + 3 C_4(\mu) \right]
\\&+& \frac{2}{9} \left( 3 C_3(\mu) + C_4(\mu) + 3 C_5(\mu) +
C_6(\mu) \right) \nonumber \,\, ,
\end{eqnarray}
and $z=m_c/m_b$. The functions $h(z,s)$ arises from the one loop contributions
of the four quark operators $O_1$,...,$O_6$ and their explicit forms can be found in
\cite{Misiak,Buras}

It is possible to parametrize  the resonance $\bar{c}c$ contribution
$Y_{reson}(s)$ in Eq.(\ref{C9efftot}) using a
Breit-Wigner shape with normalizations fixed by data which is given by \cite{AAli2}
\begin{eqnarray}
Y_{reson}(s)&=&-\frac{3}{\alpha^2_{em}}\kappa \sum_{V_i=\psi_i}
\frac{\pi \Gamma(V_i\rightarrow \ell^+
\ell^-)m_{V_i}}{q^2-m_{V_i}+i m_{V_i}
\Gamma_{V_i}} \nonumber \\
&\times & [ (3 C_1(\mu) + C_2(\mu) + 3 C_3(\mu) + C_4(\mu) + 3
C_5(\mu) + C_6(\mu))]\, .
 \label{Yresx}
\end{eqnarray}
The phenomenological parameter $\kappa$
in Eq. (\ref{Yresx}) is taken as $2.3$ so as to reproduce the correct value
of the branching ratio $BR(B\rightarrow J/\psi ~ X\rightarrow X\ell\bar{\ell})=
BR(B\rightarrow J/\psi ~ X) BR(J/\psi\rightarrow  ~ X \ell\bar{\ell})$.

Neglecting the mass of the $s$ quark, the effective short distance
Hamiltonian for the $b \rightarrow s \ell^+ \ell^-$ decay leads to
the QCD corrected matrix element:
\begin{eqnarray}\label{genmatrix}
{\cal M} &=&\frac{G_{F}\alpha}{2\sqrt{2}\pi }V_{tb}V_{ts}^{\ast }%
\Bigg\{C_{9}^{eff}(m_{b})~\bar{s}\gamma _{\mu }(1-\gamma _{5})b\,\bar{\ell}%
\gamma ^{\mu }\ell +C_{10}(m_{b})~\bar{s}\gamma _{\mu }(1-\gamma _{5})b\,\bar{%
\ell}\gamma ^{\mu }\gamma _{5}\ell  \nonumber \\
&-&2C_{7}^{eff}(m_{b})~\frac{m_{b}}{q^{2}}\bar{s}i\sigma _{\mu \nu
}q^{\nu }(1+\gamma _{5})b\,\bar{\ell}\gamma ^{\mu }\ell \Bigg\},\nonumber\\
\end{eqnarray}
where $q$ is the momentum transfer.

\subsection{The exclusive \Bpll decay in the 2HDM}
In this section we  calculate the $BR$ and the $A_{FB}$
of the \Bpll decay. In order to find these physically measurable quantities at
hadronic level, we need the following matrix elements
$<\phi(p_\phi,\varepsilon)|\bar{s}\gamma_{\mu}(1-\gamma_5)b|B(p_B)>$ and
$<\phi(p_\phi,\varepsilon)|\bar{s}i\sigma_{\mu\nu}q^\nu(1+\gamma_5)b|B(p_B)>$, which
can be parametrized  in terms of form factors. Using the parametrization of
the form factors as in  \cite{abhh}, we get the following
expression for the matrix element of   the \Bpll decay:
\bea
\label{matrixBrll}\cal{M}^{B\rightarrow\phi}&=&\frac{G_F
\alpha}{2\sqrt{2}\pi}V_{tb}V_{ts}^\ast\Bigg \{
\bar{\ell}\gamma_\mu\ell[2A\epsilon_{\mu\nu\lambda\sigma}
\varepsilon^{\ast\nu} p_\phi^\lambda p_B^\sigma +i B
\varepsilon^\ast_{\mu}-i C(p_B+p_\phi)_\mu (\varepsilon^\ast q)-i
D
(\varepsilon^\ast q)q_\mu]\nnb\\
&+& \bar{\ell}\gamma_\mu \gamma_5 \ell[2E
\epsilon_{\mu\nu\lambda\sigma}\varepsilon^{\ast\nu} p_\phi^\lambda
 p_B^\sigma +i F \varepsilon^\ast_{\mu} -i G(\varepsilon^\ast q)(p_B+p_\phi)
-i H(\varepsilon^\ast q) q_\mu]\Bigg \} \eea where
\bea A&=&C^{eff}_9\frac{V}{m_B+m_\phi}+4\frac{m_b}{q^2}C^{eff}_7 T_1,\nnb\\
B&=&(m_B+m_\phi)\Bigg( C^{eff}_9 A_1+\frac{4
m_b}{q^2}(m_B-m_\phi)C^{eff}_7
T_2\Bigg),\nnb\\
C&=&C^{eff}_9\frac{A_2}{m_B+m_\phi}+
4\frac{m_b}{q^2}C^{eff}_7\Bigg(T_2+\frac{q^2}{m_B^2-m_\phi^2}T_3\Bigg),\nnb\\
D&=&2C^{eff}_9\frac{m_\phi}{q^2}(A_3-A_0)-4C^{eff}_7\frac{m_b}{q^2} T_3,\nnb\\
E&=&C_{10} \frac{V}{m_B+m_\phi},\\
F&=&C_{10}(m_B+m_\phi)A_1,\nnb\\
G&=&C_{10}\frac{A_2}{m_B+m_\phi},\nnb\\
H&=&2C_{10}\frac{m_\phi}{q^2}(A_3-A_0),\nnb\eea Here $A_0$, $A_1$,
$A_2$, $A_3$, $V$, $T_1$, $T_2$ and $T_3$ are the relevant form
factors.

The matrix element in Eq.(\ref{matrixBrll}) leads to the following  differential
decay rate \cite{acs}
\bea \frac{d\Gamma}{ds}&=&\frac{\alpha^2 G_F^2
m_B}{2^{12} \pi^5}|V_{tb} V^*_{ts}|^2
\sqrt{\lambda_\phi}~~v~~\Delta_{\phi} \eea where \bea
\Delta_{\phi} & = & \frac{8}{3}\lambda_\phi m_B^6 s ((3-v^2)|A|^2+
2 v^2 |E|^2) + \frac{1}{r_\phi}\lambda_\phi m_B^4
\Bigg[\frac{1}{3}\lambda_\phi m_B^2
(3-v^2)|C|^2+m_B^2s^2 (1-v^2)|H|^2 \nnb \\
& + & \frac{2}{3}[(3-v^2)\, W_\phi-3 \,s (1-v^2)] Re[F~G^\ast]-2\,
s \,(1-v^2)Re[F~H^\ast] \nnb \\ &+&2 \,m_B^2 s
(1-r_\phi)(1-v^2)Re[G~H^\ast]+
\frac{2}{3}(3-v^2)W_\phi Re[B~C^\ast] \Bigg ]\nnb \\
& + & \frac{1}{3 r_\phi} m_B^2 \Bigg [ (\lambda_\phi +12 r_\phi
s)(3-v^2)|B|^2
+\lambda_\phi m_B^4 [\lambda_\phi (3-v^2)-3 s (s-2 r_\phi-2)(1-v^2)]|G|^2\nnb \\
& + & (\lambda_\phi (3-v^2)+24 r_\phi s v^2)|F|^2 \Bigg]
\label{deltaphi}.
\eea
where $s=q^2/m_B^2$, $r_\phi=m_\phi^2/m_B^2$, $v=\sqrt{1-\frac{4t^2}{s}}$, $t=m_l/m_B$,
$\lambda_{\phi}=r_\phi^2+(s-1)^2-2r_\phi(s+1)$ and $W_\phi=-1+r_\phi + s$.
Here,  $z=\cos\theta$,
where $\theta$ is the angle between the three-momentum of the
$\ell^-$ lepton and that of the $B_s$-meson in the center of mass
frame of the dileptons $\ell^+\ell^-$.

The $A_{FB}$ is another observable that can give more precise
information at hadronic level. We write its definition as given by
\begin{eqnarray}\label{genafb}
A_{FB}(s)& = & \frac{ \int^{1}_{0}dz \frac{d \Gamma }{dz} -
\int^{0}_{-1}dz \frac{d \Gamma }{dz}}{\int^{1}_{0}dz \frac{d
\Gamma }{dz}+ \int^{0}_{-1}dz \frac{d \Gamma }{dz}} \, ,
\end{eqnarray}
where $\Gamma$ is the total decay rate. The $A_{FB}$ for the \Bpll
decay is calculated to be
\bea\label{afbphi}
A_{FB}&=&\int~ds~8m_B^4\lambda_\phi v^2
s(Re[B~E^\ast]+Re[A~F^\ast])\Bigg/\int~ds \sqrt{\lambda_\phi}
~~v~~ \Delta_{\phi}.
 \eea

\section{Numerical results and discussion \label{s3}}
In this section we present the numerical analysis of the exclusive
\Bpll decay in the 2HDMs. The input parameters we used in this analysis are
as follows:
\begin{eqnarray}
& & m_{B_s} =5.28 \, GeV \, , \, m_b =4.8 \, GeV \, , \,m_c =1.4 \,
GeV \, , \, m_{\mu} =0.105 \, GeV \, ,  \nnb \\& & m_{\tau} =1.77 \, GeV \, ,
m_{\phi}=1.02 \, GeV \, , m_{H^{\pm}}=400\, GeV\, ,\, |V_{tb} V^*_{ts}|=0.04 \, , \,
\nnb
\\& &\alpha^{-1}=129 \, ,
\,G_F=1.17 \times 10^{-5}\, GeV^{-2} \,  , \,\tau_{B_s}=1.54
\times 10^{-12} \, s \,  .
\end{eqnarray}

\begin{table}
\begin{center}
\begin{tabular}{|c| c c c||c c c|}
 \hline\hline
   %after \\: \hline or \cline{col1-col2} \cline{col3-col4} ...
    && QCDSR &&&CQM&\\\hline
   & $F(0)$ & $a_F$ & $b_F$ &$F(0)$ & $a_F$ & $b_F$\\ \hline
  $A_1^{B\rar\phi}$ & $0.30$ & 0.87 & -0.06&0.59&-0.11&0.49 \\
  $A_2^{B\rar\phi}$ & $0.26$ & 1.55 & 0.51 &0.73&0.78&-0.52\\
  $V^{B\rar\phi}$ & $0.43$ & 1.75 & 0.74 &0.20&0.65&0.96\\
  $T_1^{B\rar\phi}$ & $0.35$ & 1.82 & 0.83 &0.21&0.78&0.07\\
  $T_2^{B\rar\phi}$ & $0.35$ & 0.70 & -0.32 &0.21&0.85&12.9\\
  $T_3^{B\rar\phi}$ & $0.26$ & 1.52 & 0.38 &0.18&0.62&-0.88\\
  \hline\hline
  \end{tabular}
  \end{center}
  \caption{The values of the parameters in Eq. (\ref{formrho}) for the
  various form factors of the transition
  $B\rar\phi$ calculated in the light-cone QCD sum rule approach (QCDSR) \cite{Ball}
  and using a Constituent Quark-Meson model (CQM) \cite{AD}. }
  \label{tabphi}
 \end{table}

The masses of the charged Higgs, $m_{H^\pm}$, the Yukawa couplings
($\xi_{ij}^{U,D}$) and the ratio of the vacuum expectation values
of the two Higgs doublets, $\tan\beta$, remain as free parameters
of the model. The restrictions on $m_{H^\pm}$, and $\tan\beta$ have been
 already discussed in section \ref{s2}.  For Yukawa couplings
, we use the restrictions coming from CLEO data \cite{CLEO},
\bea BR(B\rightarrow X_s \, \gamma)& = &
(3.15\pm 0.35\pm 0.32)10^{-4} \, ,\label{data}
\eea
$B^0-\bar{B}^0$ mixing \cite{Aliev4}, $\rho$ parameter
\cite{Soni}, and neutron electric-dipole moment \cite{D.Bowser},
that yields $\bar{\xi}^D_{N,ib}\sim 0$ and $\bar{\xi}^D_{N,ij}\sim
0$, where the indices $i$, $j$ denote d and s quarks, and
$\bar{\xi}^U_{N,tc}<<\bar{\xi}^U_{N,tt}$. Therefore, we take into
account only the Yukawa couplings of b and t quarks, $
\bar{\xi}^U_{N,tt}$, $\bar{\xi}^D_{N,bb}$. There is also a
restriction  on the Wilson coefficient $C_7^{eff}$ from the BR of $B\rar X_s \gamma$
in Eq.(\ref{data}) as follows \cite{Aliev4},
\bea 0.257\leq|C_7^{eff}|\leq 0.439. \eea

In our numerical calculations for  \Bpll decay, we use three
parameter fit of the light-cone QCD sum rule \cite{Ball} which can
be written in the following form
\bea\label{formrho}
F(q^2)=\frac{F(0)}{1-a_F~q^2/m_B^2+b_F(q^2/m_B^2)^2} \eea
where
the values of the parameters $F(0)$, $a_F$ and $b_F$ are given in
Table (\ref{tabphi}). This table also contains the values of the same parameters
calculated in the CQM \cite{AD}, which we use to calculate the SM predictions for
the BR in table \ref{tabBR}. The form factors $A_0$ and $A_3$ in Eq. (\ref{formrho})
can be found from the following parametrization,
\bea\label{paramet} A_0&=&A_3-\frac{T_3~q^2}{m_\phi
m_b},\nonumber\\
A_3&=&\frac{m_B+m_\phi}{2m_\phi}A_1-\frac{m_B-m_\phi}{2m_\phi}A_2.
\eea

We note that there are five possible resonances in the $c\bar{c}$ system that
can contribute to the decay under consideration and to calculate
them, we need to divide the integration region for
$q^2$ into three parts for $\ell=e,\mu $ so that we have $4
m^2_{\ell} \leq q^2 \leq (m_{\psi_1}-0.02)^2$ and
$(m_{\psi_1}+0.02)^2 \leq q^2 \leq (m_{\psi_2}-0.02)^2$ and
$(m_{\psi_2}+0.02)^2 \leq q^2 \leq (m_B-m_{\phi})^2$ , while for
$\ell=\tau$ it takes the form given by $4 m^2_{\tau} \leq q^2 \leq
(m_{\psi_2}-0.02)^2$ and $(m_{\psi_2}+0.02)^2 \leq q^2 \leq
(m_B-m_{\phi})^2$ .Here, $m_{\psi_1}$ and $m_{\psi_2}$ are the
masses of the first and the second resonances, respectively. The
SM predictions for the integrated branching ratios for $\ell=e,\mu
,\tau$ are presented in Table \ref{tabBR} for  distinct $q^2$
regions and also the total contributions.

In the following, we give the results of our calculations about the dependencies of the
differential branching ratio $(dBR/dq^2)$ and $A_{FB}(q^2)$ of the
\Bpmm decay on the four momentum transfer $q^2$, and also
dependencies of the BR and $A_{FB}$ on the model parameters .
 The results are presented by a
series of graphs, which are plotted for $\ell=\mu$ and  for the
case of the ratio
$|r_{tb}|\equiv\Bigg|\frac{\bar{\xi}^D_{N,tt}}{\bar{\xi}^D_{N,bb}}\Bigg|<1$.
We do not present the results for  $r_{tb}>1$ case, since BR for
$r_{tb}>1$ case indicates one-to-two orders of magnitude
enhancement with respect to the SM results, which seems to
conflict with the  experimental data given by BELLE collaboration
\cite{BELLEson} for the inclusive $B \rightarrow X_s \ell^+
\ell^-$ decays : \bea BR (B\rightarrow X_s \ell^+ \ell^-)=(7.1 \pm
1.6 ^{+1.4}_{-1.2})\times 10^{-6} . \eea
\begin{table}
\begin{center}
\begin{tabular}{|c|c|c|c|c|c|}
 \hline\hline
   %after \\: \hline or \cline{col1-col2} \cline{col3-col4} ...
   & {\scriptsize Short distance}   &
    $4m_l^2\leq q^2 \leq $ & $(m_{\psi_1}+0.02)^2\leq q^2 $&$(m_{\psi_2}+0.02)^2\leq q^2
    $&{\scriptsize Short+Long} \\$\ell$ & {\scriptsize contribution}&$(m_{\psi_1}-0.02)^2$&$\leq
    (m_{\psi_2}-0.02)^2$&$\leq(m_B-m_\phi)^2$& {\scriptsize distance contribution} \\
&{\scriptsize to the BR} &   & & & {\scriptsize  to the BR}\\
\hline\hline
e~~{\scriptsize (QCDSR)}&$2.01\times10^{-6}$&$1.50\times10^{-6}$&$4.75\times
10^{-7}$&$3.70\times10^{-7}$&$2.35\times10^{-6}$\\~~~~
{\scriptsize (CQM)}&$1.87\times10^{-6}$&$1.59\times10^{-6}$&$3.10\times
10^{-7}$&$1.80\times10^{-7}$&$2.08\times10^{-6}$\\
\hline
$\mu$~~{\scriptsize (QCDSR)}&$1.65\times10^{-6}$&$1.00\times10^{-6}$&$4.74\times
10^{-7}$&$3.69\times10^{-7}$&$1.91\times10^{-6}$\\~~~~{\scriptsize %%@
(CQM)}&$1.25\times10^{-6}$&$0.96\times10^{-7}$&$3.07\times
10^{-7}$&$1.08\times10^{-7}$&$1.45\times10^{-6}$\\
\hline
$\tau$~~{\scriptsize (QCDSR)}&$1.38\times10^{-7}$&$3.01\times10^{-8}$&$9.47\times
10^{-8}$&
&$1.25\times10^{-7}$\\~~~~{\scriptsize (CQM)}&$2.28\times10^{-7}$&$3.42\times10^{-8}$&$1.75\times
10^{-7}$&&$2.09\times10^{-7}$\\
\hline\hline
  \end{tabular}
  \end{center}
  \caption{The
SM predictions for the integrated branching ratios for $\ell=e,\mu
,\tau$ of the \Bpll decay for  distinct $q^2$ regions and also the
total contributions. Here, QCDSR (CQM) stands for the results calculated with the form factors
of ref. \cite{Ball} (\cite{AD}).}\label{tabBR}
\end{table}
In Fig.(\ref{BRtanbeta}), we plot the dependence of the BR of the
\Bpmm decay on $\tan\beta$ by taking $m_{H^\pm}=400 GeV $. Here the solid
(dashed) curve represents the
Model I (II) prediction for the BR and
the small dashed straight line is for the SM result.
We see that BR  decreases
with increasing values of $\tan\beta$. For $1\lsim \tan \beta \lsim 4$,
it is possible to enhance the BR $15(13)\%-2(1.5)\%$ in model I(II) compared
to its value in the SM. However the larger values of $\tan\beta$
have been favoured by the recent experimental results \cite{CERN}. We can therefore
conclude that charged Higgs boson contributions calculated in the context of the
model I and II versions of the 2HDM are not very sizable, i.e., when
$\tan \beta \gsim 5$ it almost coincides with the SM result in model I,
while in model II, its value remains slightly below the SM one.

We have the following prescription for the graphs we give in between
Figs.(\ref{dBRrkvb1a})-(\ref{AFBkbbrk1}): the regions between the solid
curves represent the $C^{eff}_7 >0$ case, the regions between the
dashed curves represent the $C^{eff}_7 <0$ case, and the small
dashed curves are for the SM predictions.

The dependence of the $dBR/dq^2$ on $q^2$ is given in Fig.(\ref{dBRrkvb1a}) in Model
III, including the long distance contributions. We see that $dBR/dq^2$ almost coincides
with the SM result for $C^{eff}_7<0$, while for $C^{eff}_7 >0$ cases it is considerably enhanced.

The dependence of the BR on one of the
free parameters of Model III, $\bar{\xi}^D_{N,bb}/m_b$ is shown in Fig.(\ref{BRkbbrk1}).
It is seen from this figure that BR is not very much sensitive to
$\bar{\xi}^D_{N,bb}/m_b$, especially for large values of this parameter. We note that
the SM
and model III average results for BR when $C^{eff}_7 >0$ ($C^{eff}_7 <0$) are
$1.91 \times 10^{-6}$ and $4.00\times 10^{-6}$ ($1.91\times 10^{-6}$).

The dependence of the BR on the charged Higgs mass $m_{H^\pm}$ is
presented in Fig.(\ref{BRmHrk1}). The BR decreases as $m_{H^\pm}$ increases, except the $C^{eff}_7
<0$ case, which is insensitive to $m_{H^\pm}$ and very close
to the SM prediction in magnitude.

We present the dependence of the $A_{FB}(q^2)$ on $q^2$ for the
\Bpmm decay in Fig.(\ref{dAFBrkvb1a}). Here, the
$A_{FB}(q^2)$ is enhanced for $C^{eff}_7>0$ while it almost coincides with the SM prediction for
$C^{eff}_7 >0$.

In Fig.(\ref{AFBkbbrk1}), the dependence of
the $A_{FB}$ on $\bar{\xi}^D_{N,bb}/m_b$ is represented. We observe that $A_{FB}$ is not very
sensitive to
$\bar{\xi}^D_{N,bb}/m_b$ , especially for large values of this parameter.
The SM and model III average results for $A_{FB}$ when $C^{eff}_7 >0$ ($C^{eff}_7 <0$)
are $-0.23$ and $-0.31$ ($-0.21$).

In conclusion, we have investigated the physical observables, BR and $A_{FB}$
related to the exclusive \Bpll decay in the model I, II and III versions of the
2HDM. We have found that these observables are highly sensitive to new physics and hence
provide powerful probe of the SM.

%\listoftables \listoffigures

\newpage
\begin{figure}[thb]
\vskip 0truein \centering \epsfxsize=3.8in
\leavevmode\epsffile{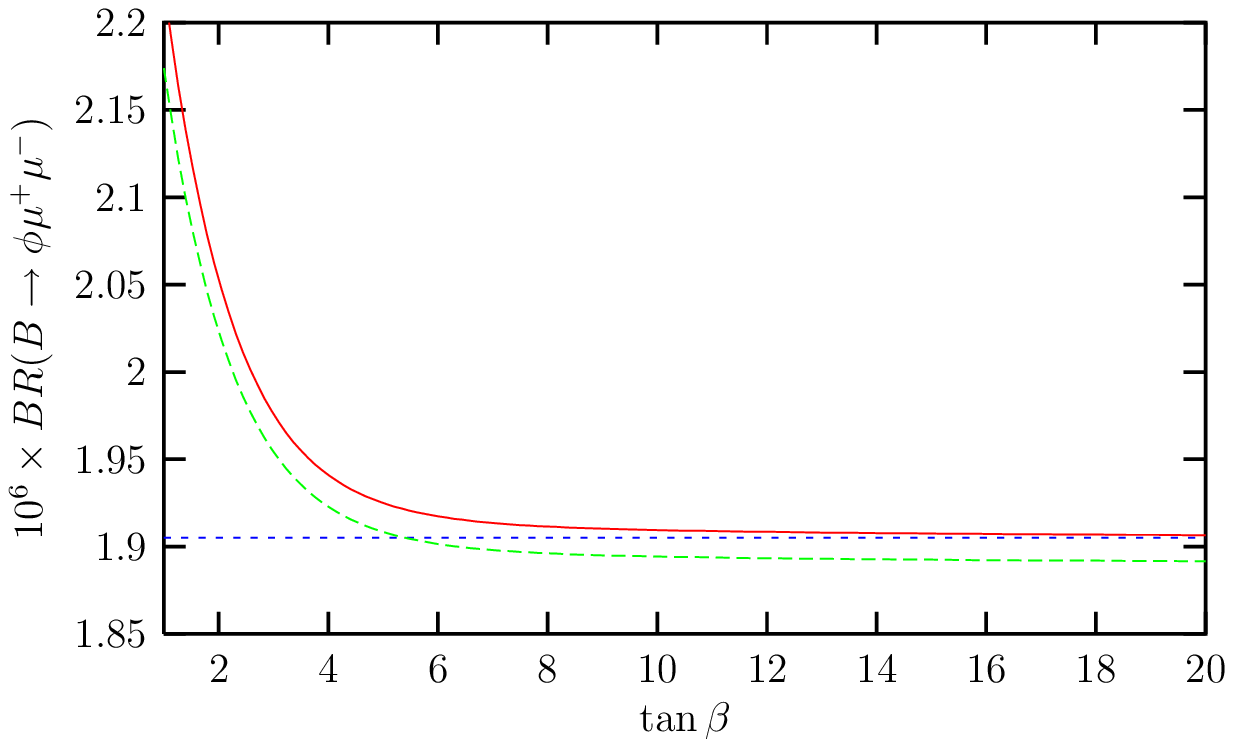} \vskip 0truein \caption[]{The
dependence of the BR of the \Bpmm decay on $\tan\beta$. Here the
solid (dashed) curve represents the Model I (II) contribution
while the small dashed straight line is for the SM case.}
\label{BRtanbeta}
\end{figure}
\begin{figure}[bht]
\vskip 0truein \centering \epsfxsize=3.8in\leavevmode
\epsffile{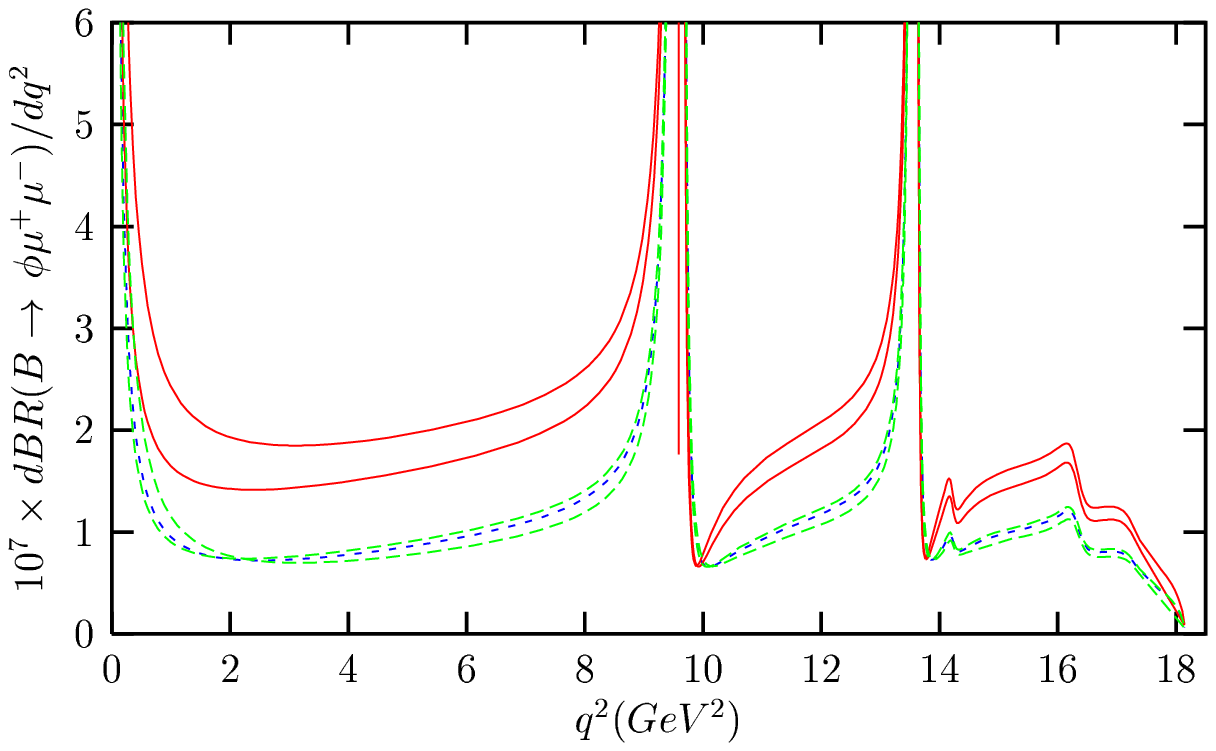} \vskip 0truein \caption{ The dependence
of the $dBR/dq^2$ on $q^2$. Here the region between the solid
curves represents the $dBR/dq^2$ for $C^{eff}_7 >0$, while the one
between the dashed curves is  for $C^{eff}_7 <0$. The SM
prediction is represented by the small dashed curve. Here we take
$\bar{\xi}_{N,bb}^{D}=40 \, m_b$. } \label{dBRrkvb1a}
\end{figure}
\begin{figure}[htb]
\vskip 0truein \centering \epsfxsize=3.8in
\leavevmode\epsffile{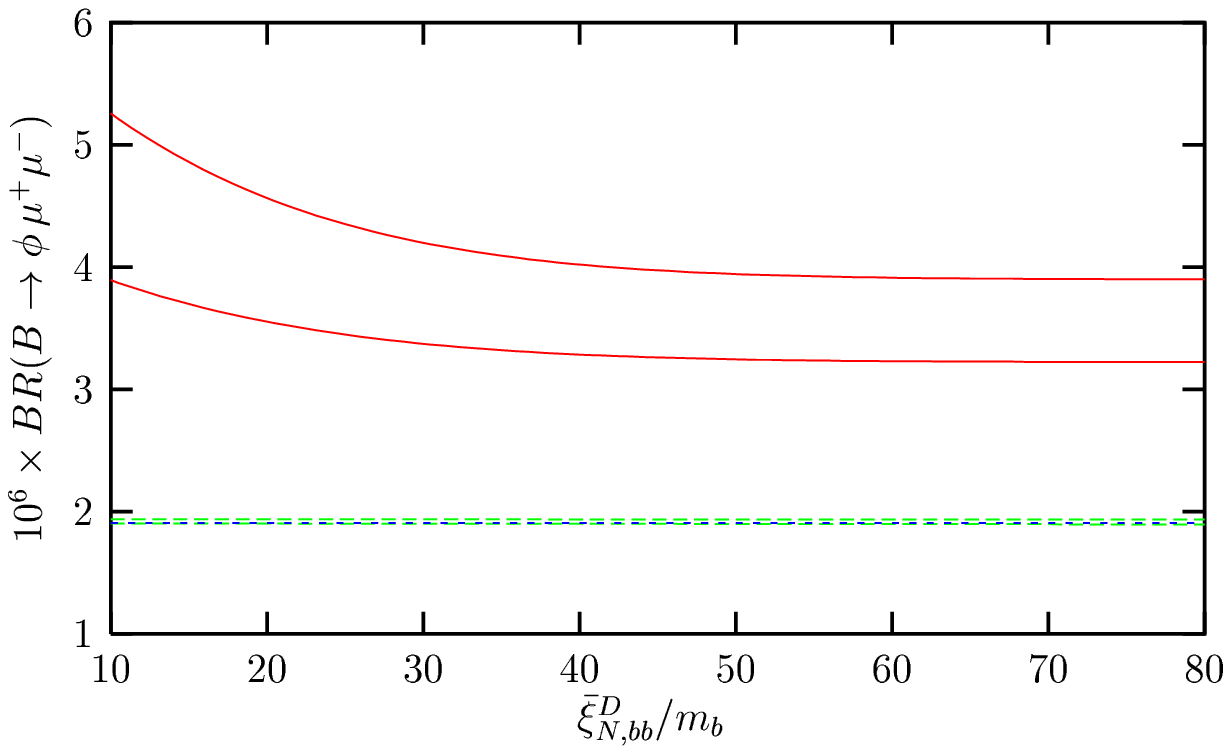} \vskip 0truein \caption{The
dependence of the BR on $\bar{\xi}^D_{N,bb}/m_b$.
Here $BR$ is restricted in the
region between solid (dashed) curves for $C^{eff}_7 >0$ ($C^{eff}_7 <0$). Small
dashed straight line represents the SM prediction.}\label{BRkbbrk1}
\end{figure}
\begin{figure}[htb]
\vskip 0truein \centering \epsfxsize=3.8in
\leavevmode\epsffile{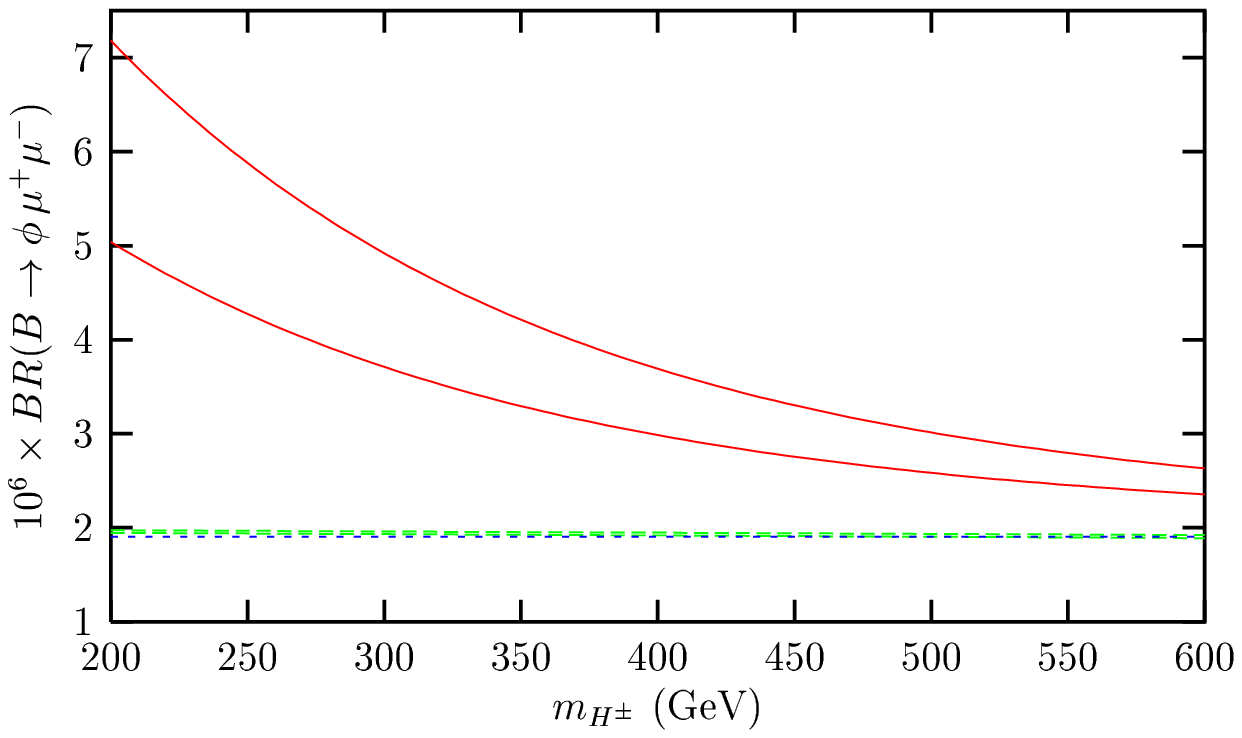} \vskip 0truein \caption{The
dependence of the BR on  $m_{H^\pm}$ for
$\bar{\xi}_{N,bb}^{D}=40\, m_b$. Here $BR$ is restricted in the
region between solid (dashed) curves for $C^{eff}_7 >0$
($C^{eff}_7 <0$). Small dashed straight line represents the SM
prediction. } \label{BRmHrk1}
\end{figure}
\begin{figure}[htb]
\vskip 0truein \centering \epsfxsize=3.8in
\leavevmode\epsffile{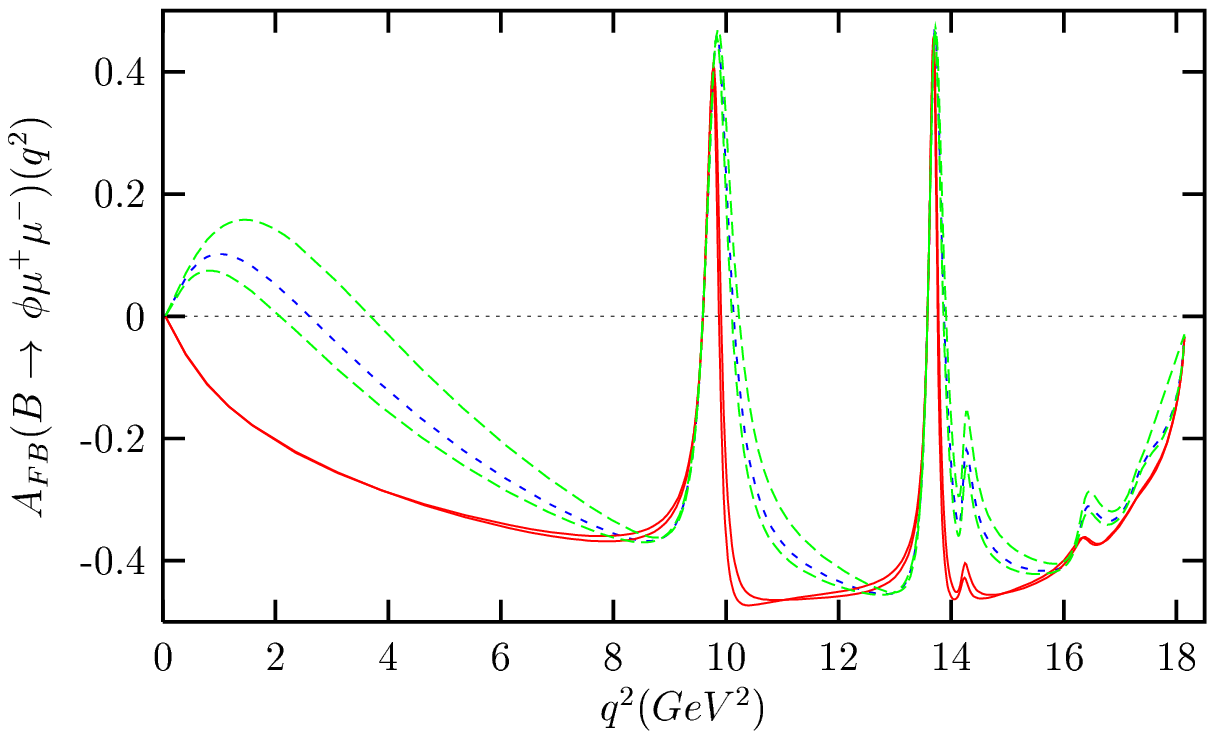} \vskip 0truein \caption{ The
dependence of the $A_{FB}(q^2)$ on $q^2$. Here the region between
the solid  curves represents the $A_{FB}(q^2)$ for $C^{eff}_7 >0$
, while the one between the dashed curves is for the $A_{FB}(q^2)$
for $C^{eff}_7 <0$. The SM prediction is represented by the small
dashed curve. Here we take $\bar{\xi}_{N,bb}^{D}=40 \, m_b$.}
\label{dAFBrkvb1a}
\end{figure}
\begin{figure}[htb]
\vskip 0truein \centering \epsfxsize=3.8in
\leavevmode\epsffile{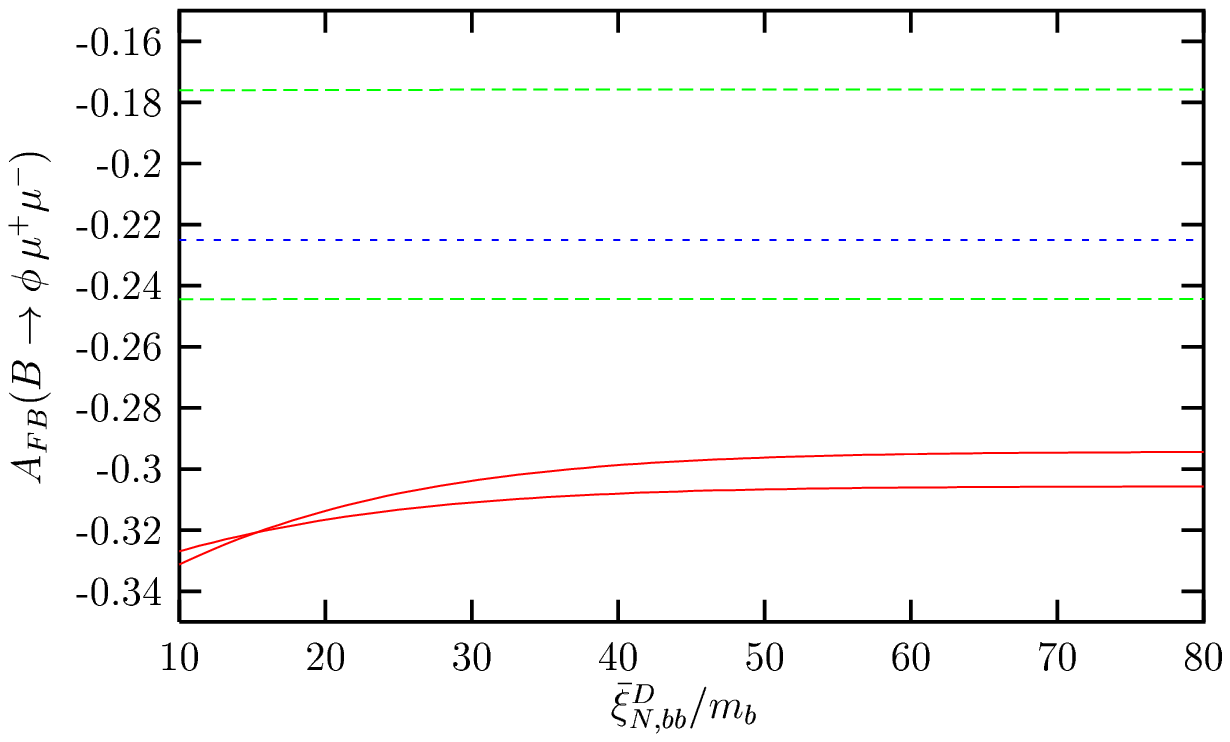} \vskip 0truein \caption{The
dependence of the $A_{FB}$ on $\bar{\xi}^D_{N,bb}$. Here $A_{FB}$ is
restricted in the region
between solid (dashed) curves for $C^{eff}_7 >0$ ($C^{eff}_7 <0$).
Small dashed straight line represents the SM prediction.}
\label{AFBkbbrk1}
\end{figure}
\end{document}